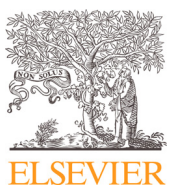

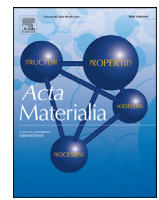

Full length article

# Crystallization of GeO₂ thin films into α-quartz: from spherulites to single crystals

Silang Zhou, Jordi Antoja-Lleonart, Pavan Nukala, Václav Ocelík, Nick R. Lutjes, Beatriz Noheda*

*Zernike Institute for Advanced Materials, University of Groningen, Nijenborgh 4, 9747AG Groningen, the Netherlands*

ABSTRACT

Piezoelectric quartz (SiO₂) crystals are widely used in industry as oscillators. As a natural mineral, quartz and its relevant silicates are also of interest in geoscience and mineralogy. However, the nucleation and growth of quartz crystals are difficult to control and not fully understood. Here we report successful solid-state crystallization of thin film of amorphous GeO₂ into quartz on various substrates, including Al₂O₃, MgAl₂O₄, MgO, LaAlO₃ and SrTiO₃. At relatively low annealing temperatures, the crystallization process is spherulitic: with fibers growing radially from the nucleation centers and the crystal lattice rotating along the growth direction with a linear dependence between the rotation angle and the distance to the core. For increasingly higher annealing temperatures, quartz crystals begin to form. The edges of the sample play an important role in facilitating nucleation followed by growth sweeping inward until the whole film is crystallized. Control of the growth allows single crystalline quartz to be synthesized, with crystal sizes of hundreds of microns achieved on sapphire substrates, which is promising for further piezoelectric applications. Our study reveals the complexity of the nucleation and growth process of quartz and provides insight for further studies.

## 1. Introduction

Quartz (SiO₂) is one of the most abundant minerals in the earth's crust. It has been widely used in electronics as the key element in oscillator circuits. These oscillators use the mechanical resonance of a piezoelectric material to generate signals with a precise frequency, which requires high crystallinity of the piezoelectric. Quartz oscillators are found in computers, cellphones, wristwatches, clocks and radios [1]. The standard method of producing high quality quartz crystals is hydrothermal growth, followed by mechanical cutting and etching [2]. However, the size of the crystals made by this top-down method are limited to tens of micrometers, which in turn limits the frequencies to hundreds of MHz and it does not meet the high frequency current trends of information and communication technologies or the miniaturization trends in the microelectronic industry [3,4].

Recently, several reports have studied the synthesis of submicron-sized quartz, such as epitaxial quartz thin films on silicon by soft chemistry methods [5,6], homoepitaxy of quartz on quartz by ion beam implantation [7,8], hydrothermal growth of nano-quartz [9,10], as well as solvothermal synthesis [11] and microemulsion-mediated synthesis of quartz [12]. The challenge lies in the strong tendency of amorphization of silicon dioxide and the difficulty to control the nucleation and growth processes. It has been proved that the rigid network of SiO₄ tetrahedra makes the solid state transformation of silica into quartz impossible unless some catalyst ions, so-called 'network modifiers' or 'melting agents', are added [8,13]. These melting agents are usually alkaline metals or alkaline earth metals, well known in the glass industry to lower the melting points of silicates to a more accessible temperature. The network modifiers can form bonds with the oxygen atoms, breaking the connectivity between the SiO₄ tetrahedra and making the network less rigid. This leads to significant lowering of the melting points of the silicate and to an increased nucleation rate [14].

We have focused our study on quartz-type GeO₂ [15,16]. At ambient conditions, the stable phase of GeO₂ is rutile and the quartz phase is meta-stable [16,17]. However, when pure GeO₂ is cooled down slowly from the melt, the quartz phase is formed instead and bulk single crystals of GeO₂ quartz can be stabilized up to 1000 °C [18], due to the extremely sluggish transformation from quartz to rutile [16]. Moreover, the melting point of GeO₂ is hundreds of de-grees lower than that of $SiO_2$, due to the weaker bonds of Ge-O compared with those of Si-O [16]. The less rigid network of $GeO_2$ makes the solid-state crystallization possible without using melting agents.

* Corresponding author.
  *E-mail address:* b.noheda@rug.nl (B. Noheda).

We report successful solid-state crystallization of thin amorphous films of pure $GeO_2$ into the quartz structure by means of controlling the crystallization temperature. We have found that, at low temperature, spherulitic crystallization is promoted due to high occurance of lattice defects [19]. With increasing crystallization temperature, columnar and dendritic growth take place and, finally at high enough temperatures, single crystals form.

Spherulites are made of fibers that grow radially, starting from a nucleation core, conferring the typical spherical (or circular, in two dimensions) shape to the spherulites. Spherulites are formed by non-crystallographic branching, where the parent and daughter fibers do not share the same crystal orientation, with misorientation angles that can range between 0-15° [19]. Using local crystal orientation inspection, we have characterized the lattice rotation along the fibers' growth directions. The formation of spherulites from the liquid phase requires high crystallization driving force, usually provided by high supersaturation or supercooling. Spherulites can be found in different kinds of materials, such as polymers [20], small molecules [21] or inorganic minerals [22]. In nature, spherulitic forms of quartz, such as chalcedony [23], and other silicates can be found in volcanic rocks. However, to our knowledge, spherulitic crystallization of pure quartz from amorphous solid has never been discussed in the literature. Most of the studies on spherulites are focused on polymers, which often have complex structures with folded chains. Our simple system may shed light upon the spherulitic growth process and its control. The single crystals promoted at high annealing temperatures can be as large as several hundreds of micrometers, which is promising for further piezoelectric applications.

## 2. Experimental

Thin amorphous films of $GeO_2$, with thicknesses of about 100 nm, were deposited by pulsed laser deposition (PLD) using a 248 nm KrF laser (Lambda Physik COMPex Pro 205). The target was made from cold pressing of $GeO_2$ (Alfa Aesar, 99.9999 %) powders. In some cases, a $TiO_2$ buffer layer with thickness of about 18nm was deposited. For that purpose, pellets were made using $TiO_2$ powders (Alfa Aesar, 99.6 %). The substrates used were: Sapphire $Al_2O_3$ (0001), MgO (110), Spinel $MgAl_2O_4$ (110), $SrTiO_3$ (STO) (001), Nb-doped $SrTiO_3$ (Nb-STO) (001), $LaAlO_3$ (LAO) (001) (Crys-Tec GmbH). Since the nucleation was found to happen at the edge of the sample, the size of the substrate influences the process. In our experiments, the size of the LAO, STO, Nb-STO substrates were $5 \times 5$ mm$^2$; $MgAl_2O_4$, MgO and $Al_2O_3$ were mechanically cut from $10 \times 10$ mm$^2$ piece to be also approximately $5 \times 5$ mm$^2$. After cutting, the substrates are cleaned in an ultrasonic bath with ethanol for several minutes to eliminate the cutting debris. Thermal treatment of the substrates before deposition did not lead to different crystallization patterns, so no treatment was performed. The substrates were glued with silver paste on a heating plate, which was heated from the back with a laser (DILAS Compact- Evolution, wavelength 808 nm). A pyrometer was used to measure the temperature of the center of the plate also from the back. Thus, the temperature of the substrates is likely to be somewhat lower than the reading of the pyrometer. We have found that the temperature inhomogeneity caused by the laser heating affects the crystallization of the sample: if the laser is moved closer to one of the substrate edges, that edge will start to crystallize (and melt if the temperature is high enough) earlier than the rest (Fig. S1). Far from being a disadvantage, this selective edge growth promotes the formation of larger areas of single crystals.

The deposition parameters of the $GeO_2$ ($TiO_2$) growth were optimized at a target-substrate distance of 46 mm (52 mm), a fluence between 2.5 -3.5 J/cm$^2$ (~2 J/cm$^2$), deposition temperature of 600 °C (600 °C), oxygen pressure of 0.1 mbar (0.1 mbar) and a laser repetition rate of 5 Hz (1 Hz) for 12 (30) minutes After the deposition, the $GeO_2$ thin films are amorphous and annealing is necessary for crystallization. The films were heated to the targeted annealing temperature with a ramp rate 25 °C/min. The annealing conditions were: a temperature of 830 °C, 880 °C or 930 °C (as described later), oxygen pressure of 200 mbar for 30 minutes. Subsequently, the films were cooled down to room temperature with a rate -5 °C/min.

The local crystallinity and crystal orientations were characterized by Electron Back-Scatter Diffraction (EBSD), performed on a FEI Nova NanoSEM 650 scanning electron microscope. The sample was pre-tilted on a 71° holder. Thin films grown on Nb-STO (conducting) substrates were observed in high vacuum with a standard secondary electron detector. Films grown on the other (insulating) substrates, were measured in 0.5 mbar of chamber pressure with a Low Vacuum Detector to minimize charging effects. EBSD data was collected in both cases using an EDAX EBSD system equipped with a Hikari CCD camera for recording Kikuchi patterns, with speed between 5-20 p/s and distances between neighbor points from 0.15 to 1 μm, at 20 keV of acceleration voltage and about 10 nA of current in the primary electron beam. EDAX OIM Analysis 8.1 and Matlab® based toolbox MTEX [24] software were used for EBSD data analysis. Inverse Pole Fig. (IPF) and Image Quality (IQ) maps were plotted along three axes [100], [010], [001] in Cartesian coordinates, where [001] is the out-of-plane direction of the sample and [100] and [010] are in-plane directions, [100] direction being horizontal on the maps. The average crystallinity was checked by X-Ray Diffraction (Panalytical X'Pert, CuK$_\alpha$ radiation) and the morphology was imaged by Atomic Force Microscope (AFM) (Bruker Dimension Icon). Local structures were further characterized by atomic resolution High Angle Annular Dark Field Scanning Transmission Electron Microscopy (HAADF-STEM) measurements using a Themis-G, double corrected electron microscope at 300 kV. Cross-sectional lamellae for STEM analysis were prepared via Focused Ion Beam (FIB, Helios 660) based procedure.

## 3. Results

### 3.1. Crystallization of GeO2 on Al2O3(0001) substrates

Fig. 1 shows the crystallization of $GeO_2$ into quartz on sapphire ($Al_2O_3$) substrates at different annealing temperatures. Films grown on these substrates are only partially crystallized and a significant volume of the material remains amorphous. When the films are annealed at 830 °C, dense and compact spherulites with diameter of around 50-100 μm are formed, as shown in Fig. 1(a)-(b). As mentioned in the introduction, these spherulites are spherical (or circular) structures that consist of fibers growing radially and quasi-isotropically from a single nucleation point. The different colors in the Inverse Pole Figure (IPF) maps indicate the different orientations of the fibers. The black lines in the maps of Fig. 1(a) (and in the rest of this work) denote the boundaries between neighboring fibers that present differences in the orientation angles larger than 3°. We denote these regions delimited by the solid lines as "grains" for the sake of the discussion. However, each such spherulitic grain is composed of many fibers as shown in Fig. S2. Within one grain, changes in orientation are visible. It can be observed that, while in the two smaller spherulites of Fig. 1(a) and (c), most of the fibers belong to the same grain (most fibers are misoriented by less than 3° with respect to their neighbors), for the big spherulites, which have nucleated from a contamination particle, many more grains are present. Statistical analysis of the misorientation angle distributions of the EBSD maps (see Fig. S3) also shows that almost 80 % of the observed misorientation angles between neighboring grains in the same spherulite are within 5°. This small angle branching is the characteristic behavior of spherulitic growth [19].

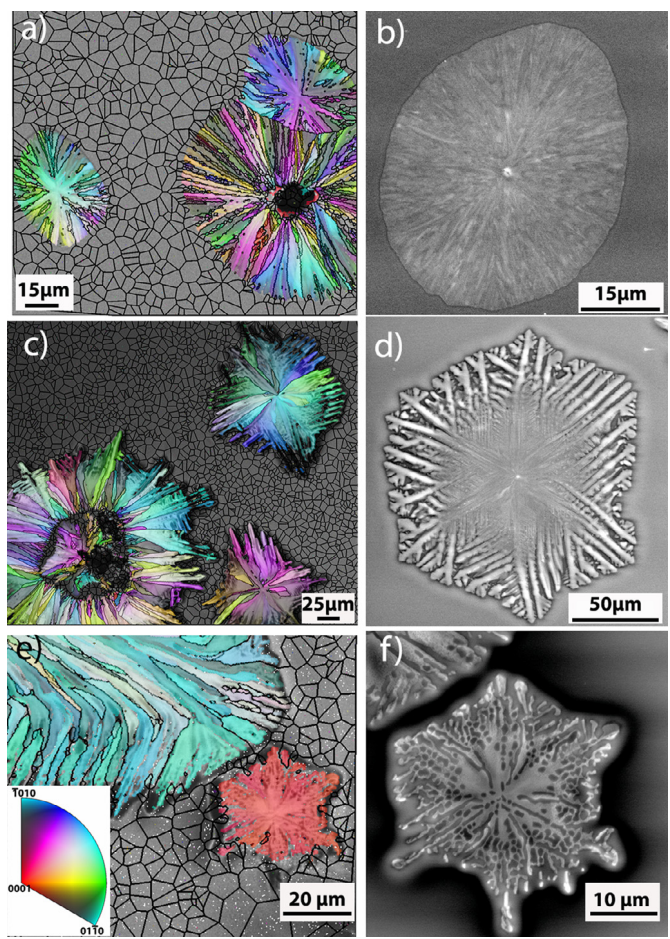

**Fig. 1.** EBSD (left) and SEM (right) images of quartz crystals on Al$_2$O$_3$ substrates crystallized with different annealing temperature, showing different crystallization features: (a-b) The films annealed at 830 °C show typical spherulites with dense radially-grown fibrils. (c-d) films annealed at 880 °C show spherulites with dendritic growth at the spherulite edge. (e-f) thin films annealed at 930 °C show single crystalline growth. (a), (c), (e) images are made by overlaying the Inverse Pole Figure (IPF) color map and the Image Quality (IQ) b&w map viewed along the out-of-plane [001] direction of the sample, in which the black lines indicate the grain boundaries with misorientation larger than 3°. (e) and (f) are taken at the same sample area. Inset in (e) shows the color legend for all three IPFs.

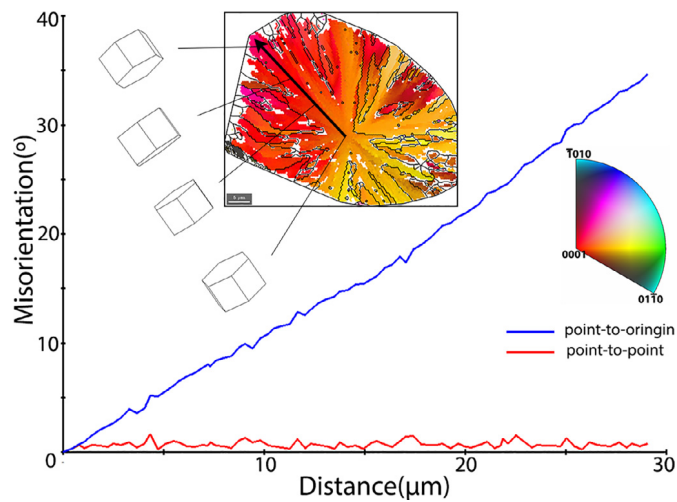

**Fig. 2.** Lattice rotation inside one fiber of a spherulite formed during crystallization at 830 °C on Al$_2$O$_3$ substrates. Lattice rotation is expressed in the form of point-to-origin crystal misorientation angles as a function of distance along the fiber starting from the center towards the edge of the spherulite. The inset represents the [110] IPF map of the spherulite, with the black arrow indicating the measurement direction. The local rotation along the fiber is visualized by means of quartz crystal shapes, viewed along [001] direction, at four different points along the black arrow.

The morphology of such spherulites is also characterized by SEM (Fig. 1(b)) and AFM (see Fig. S2). The AFM images reveal a protruding core of about 500 nm in diameter at the center of the spherulite, serving as a nucleation center. Fibrils grow radially from the core. It is also visible that the spherulites are lower in height compared to the amorphous surroundings, as expected due to densification upon crystallization. We have found that contamination particles can facilitate nucleation by serving as core for the spherulites (see Fig. S4). Generally speaking, they have irregular shapes and often have larger sizes, about several μm to tens of μm, than the cores discussed above. The substrate cutting process may generate these particles, but the possibility that they come from the PLD plume during the deposition cannot be ruled out [25,26].

As shown in Fig. 1(c) and (d), increasing the annealing temperature promotes a more anisotropic growth, with fingering patterns at their growth fronts and the shape of some crystals reflecting the symmetry of quartz (P3$_2$21 or P3$_1$21) [27]. It can be observed how the number of grains decreases dramatically compared to those at the lower temperature growth, leaving only few of them with exaggerated aspect ratio. As shown from the SEM image (Fig. 1(d)) and AFM (see Fig. S2), the center of the spherulite is flat with a core in the middle, and it develops a hexagonal shape, reflecting the symmetry of the (0001) plane of quartz. At some distance from the core, it branches into bundles of fine fibers forming a rougher structure. Fractal growth of daughter fibrils protruding also at specific angles with respect to the parent ones fills the space.

If the annealing temperature is increased further to 930 °C, single-crystal quartz domains are formed as shown on the right side of the EBSD map in Fig. 1(e) and as a magnified SEM image in 1(f). The homogeneous red color of the crystal in the [001] IPF map indicates that the c-axis of the crystal points out of surface plane, and its hexagonal shape reflecting the symmetry of the quartz in that orientation. AFM scans (Fig. S2) show no fibrous growth, suggesting a different crystallization mechanism. Moreover, some bumps about 200 nm in diameter which give dark contrast in the SEM image appear, which suggests they may be porous amorphous material which has not crystallized yet.

As mentioned above, Fig. 1(a), (c) shows that in each grain which is composed of fibers, whose local crystal direction is changed gradually as the fiber grows. A quantitative analysis reveals that the quartz lattice rotates linearly along the fiber growth direction, as shown in the example of Fig. 2. The point-to-origin crystal rotation angle increases linearly along the fiber with an average slope of 1.25 °/μm. In addition, the point-to-point misorientation angle value along the same line does not exceed 2° (distance between scanning points is 0.4 μm), which explains why the selected fiber is contained in one grain, despite the large misorientation that exists between its beginning and its end (35°).

This kind of rotating lattice has been previously observed in similar amorphous to crystalline solid transformation (e.g. in Fe$_2$O$_3$ [28] or Sb$_{3.6}$Te [29] by in situ electron beam annealing in TEM, or on Sb$_2$S$_3$ [30] by laser surface annealing). Savytskii et al., suggest that the lattice rotation is accompanied by re-arrangements of dislocations and conclude that unpaired dislocations, caused by the stress generated from the volume change during crystallization, are the origin of the continuous lattice rotation [30]. Single quartz grain rotation during its growth from thin amorphous layers has deserved a separate study [31].

Interestingly, the typical crystal rotation gradient inside a fiber of the spherulites annealed at 830 °C is about 0.5-1.5 °/μm, while for those annealed at 880 °C, the rotation rate decreased to about 0.25-0.5 °/μm, being absent for the single crystals formed during annealing at 930 °C. The decreased lattice rotation with increasing annealing temperature may be associated to a smaller stress due to the smaller volume change taking place at higher temperature, accompanied by a higher dislocations mobility. Since reported spherulitic lattice rotation happens in solid-to-solid, amorphous-to-crystalline transformation, we propose that the single crystal domain formed during annealing at 930 °C are nucleated from, or close to the liquid phase.

Besides the spherulites and single crystals that are grown from a single nucleation point described above, the largest crystallized areas belong to crystals that nucleate from the edge of the sample and sweep inwards (see green area in Fig. 1(e)). As shown in the Fig. S5, a line of half spherulites are formed at the edge of the sample, which shows the strong preference for nucleation at the sample edge. Some of their fibers grow beyond the circular peripheries of the spherulites, resulting in a continuous spherulites area occupying most of the crystallized area. From now on, we will call the spherulites grown from single nucleation point, as those shown in Fig. 1, single-nucleus spherulites, to distinguish them from the continuous spherulites regions, which are an assembly of fibers grown from multiple nuclei and usually starting from the edge of the sample.

Fig. 3 shows these larger continuous spherulites of a film annealed at 830 °C. At this relatively low temperature, these crystalline areas also show spherulitic growth with needle-like fibers growing from the sample edge and forming a circular and smooth growth front. Lattice rotation can also be clearly observed in some of the grains as indicated by their color gradients, while for other grains there is almost no lattice orientation change. Interestingly, two types of modulations of the surface are detected by AFM in a few locations marked on Fig. 3(a): large concentric surface waves and a multitude of tiny trenches coexist in those regions as shown in Fig. 3(d-g). The quasi-periodic waves that modulate the surface morphology and the interference between two modulation fronts is clearly captured in Fig. 3(d-g).

In these images, the continuous concentric shape of one growing front is deformed by a neighboring one that remains concentric (see Fig. 3(d-g)), indicating that the growth fronts were not formed at the same time. This also seems to indicate that the waves, giving rise to the large concentric steps, are not formed during the crystallization process, but during the cooling down process, when internal stress accumulated due to a difference in thermal expansion coefficient between GeO$_2$ quartz and the substrate. The stress is relieved subsequently by local slip in two slip planes with local maximal shear stress components. On the other side, the system of dense trenches or tiny surface steps observed in Fig. 3(d-g) is probably formed immediately after or during crystallization. The fact that these trenches are detected in fiber areas without lattice rotation suggests that the strain that arose during crystallization in these areas is not used to locally bend the crystal, but instead it is relieved by dislocation movement along appropriate slip planes, forming the tiny steps at the surface.

Fig. 4 shows one of the continuous spherulites for the films annealed at 880 °C. In this area, the crystals show preferred orientation with (10$\bar{1}$0) out-of-plane and [0001] direction parallel to [001] sample direction. The pole Figure in Fig. 3(d-e) attests to the strong texture of the film. This orientation is the same as that of epitaxial thin films of quartz (SiO$_2$) grown on silicon [5], which was attributed it to the lattice match with the silicon substrate. In our study, the substrate is sapphire (0001), which has a hexagonal in-plane lattice with a= 4.758 Å, c= 12.988 Å [32] while the lattice parameter for bulk GeO$_2$ is a= 4.984 Å, c= 5.647 Å [15]. This does not offer a good lattice matching, being the smallest lattice mismatch 4.54 % along the [11$\bar{2}$0] direction. However, while EBSD analysis provides information of the crystal orientation locally, XRD analysis provides an overview of the orientation for the whole sample. As shown in the 2θ − ω scan in Fig. S7, indeed the (10$\bar{1}$0) peak shows up, which confirms the EBSD observation in Fig. 4. However, the presence of the (10$\bar{1}$1) and (11$\bar{2}$1) peaks in the XRD diffraction patterns (see Fig. S7), indicates that other areas in this sample have crystallized with texture like shown in Fig. 4 but with different orientations.

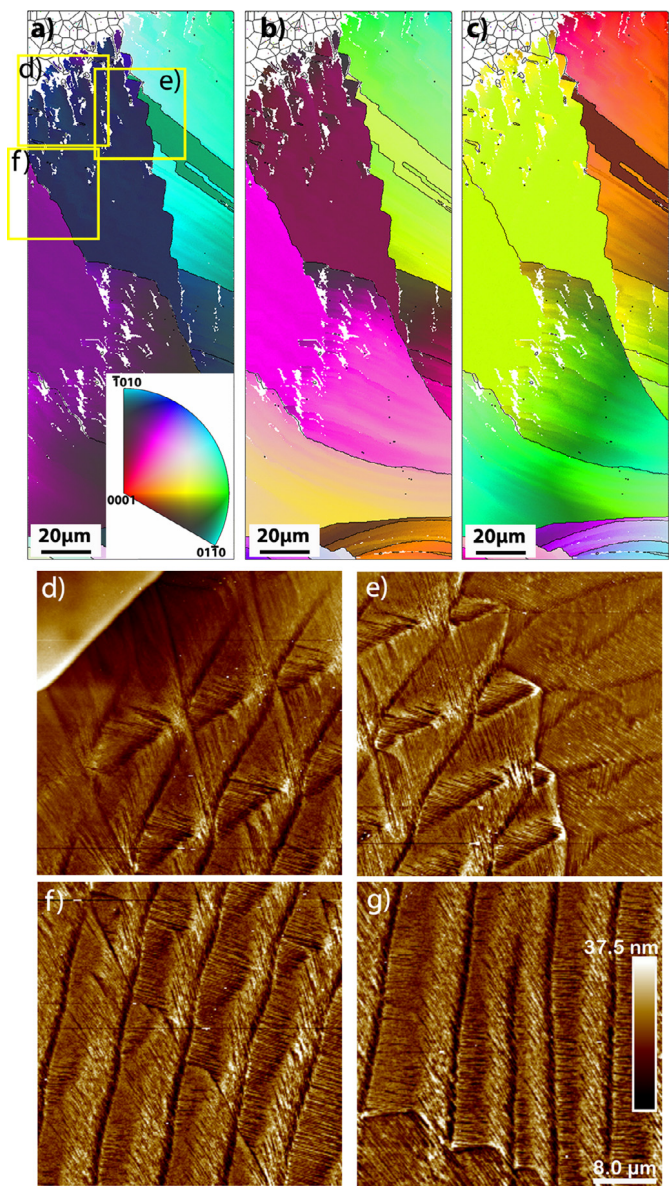

Fig. 3. Continuous spherulites grown from the sample edge on Al$_2$O$_3$ substrate annealed at 830 °C. (a-c) IPF map viewed along [100], [010] and [001] directions, respectively. (scans were taken from the area in the black rectangle in Fig. S6). (d-f) AFM scans at the place indicated by the corresponding yellow squares in a) show fibrous growth with periodic modulations in height, reflecting in each case, two superimposed growth wave fronts. (g) AFM scan at a different location in Fig. S6. The scale bar at (g) serves for all AFM scans.

From the observation of grain boundaries (solid lines in the Fig. 4(a)), it is obvious that the initial stages of the growth close to the edge nuclei, the growth is spherulitic, with the structures composed of many fibers; while at the growth front, the grains are differently oriented resembling columnar growth. The grain boundaries in Fig. 4(a) indicate that these columns are just continuation of the fibers, but in different direction. The majority of these columns have their longitudinal, e.g., growing axis, well aligned with [0001] crystal direction.

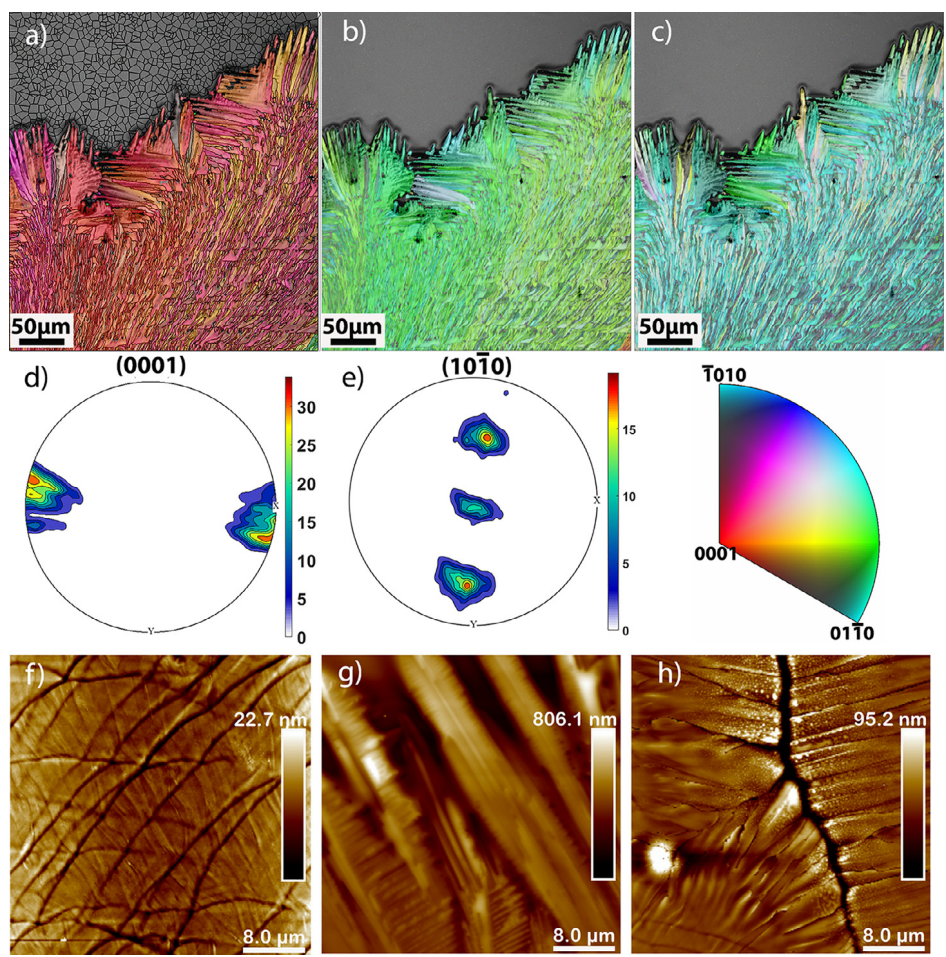

**Fig. 4.** Quartz crystals grown from the sample edge on $Al_2O_3$ substrates annealed at 880 °C show spherulitic growth at the beginning (closer to the bottom edge) but with columnar-dentritic growth at the growth front. (a-c) [100], [010] and [001] IPF maps, respectively, with near uniform color suggest the preferred $(10\bar{1}0)$ out-of-plane orientation. (d-e) Pole Figures of this area showing the strong texture. (f) AFM image of the spherulitic growth with the waves imposed on the fibers. (g) AFM image of the dendritic growth front and the depletion region around it. (h) AFM image of two growth fronts meeting with each other and the boundary between them.

Fig. 4(f) shows the morphology of a typical area of the spherulitic part, where fibers can be clearly observed. These fibers are thicker than those of film annealed at 830 °C and two sets of waves intercept with them, but they are less organized compared to the waves in Fig. 3(d-g). As mentioned above, at the growth front, oriented crystals with dendrite-like structures are found, similar to those in the single-nucleus spherulites. AFM scans (Fig. 4g)) show that the crystalline dendrites can be as high as 800 nm. Considering the thin film is only about 100 nm thick, a large mass of the material has migrated to the growth front to form these fingers, resulting in the depletion region around the fingers, which may prevent further growth.

At the growth front, lattice rotation is also observed at this higher temperature with a rate of about 0.2-0.6 °/μm, which is similar to that of the single-nucleus spherulites described earlier. On the other hand, inside the spherulitic region, the surface modulations modify the morphology as shown in Fig. 4(f), giving rise to short and curved fibers, which makes the measurements more difficult than at the growth fronts. Further analysis shows that for some fibers, there is a linear lattice rotation while for other fibers, there is lattice change along the growth direction without a clear relationship.

Fig. 5 shows the crystallization of $GeO_2$ under a still higher annealing temperature of 930 °C. Interestingly, large pieces of single crystals, which are about 100-200 μm long and 100 μm wide coexist with the spherulites. It is easy to differentiate the single crystals from the spherulites under the SEM, since the single crystals give a uniform contrast while for the spherulites, fibrous morphology can be clearly seen. These morphologies are also observed in the AFM images of Fig. 5 (b-c) and it is clear that the single crystals are flat with roughness, $R_q \sim 0.4nm$, which makes them suitable for fabricating piezoelectric devices.

We have studied the crystals in the yellow square in Fig. 5 by EBSD and the result is shown in Fig. 5 (d-e). For the spherulites, lattice rotation still occurs in some areas, but the general linear relationship, shown in Fig. 2, does not hold anymore. Four single crystals are captured in the image, and the homogenous color indicates there is no lattice rotation in them. Single crystals with orientation close to $[0\bar{1}12]$ and [0001] out-of-plane appear most often, with Fig. 5 (d-e) as a typical example. To have a better overview, we have numbered the single crystals in one sample and checked their orientations by EBSD, and the result is summarized in Fig. S8. Generally speaking, in terms of the probability of occurrence, about 1/3 of the crystals orient with $[0\bar{1}12]$ out-of-plane, 1/3 with [0001] out-of-plane, and crystals with a variety of other orientations compose the rest 1/3.

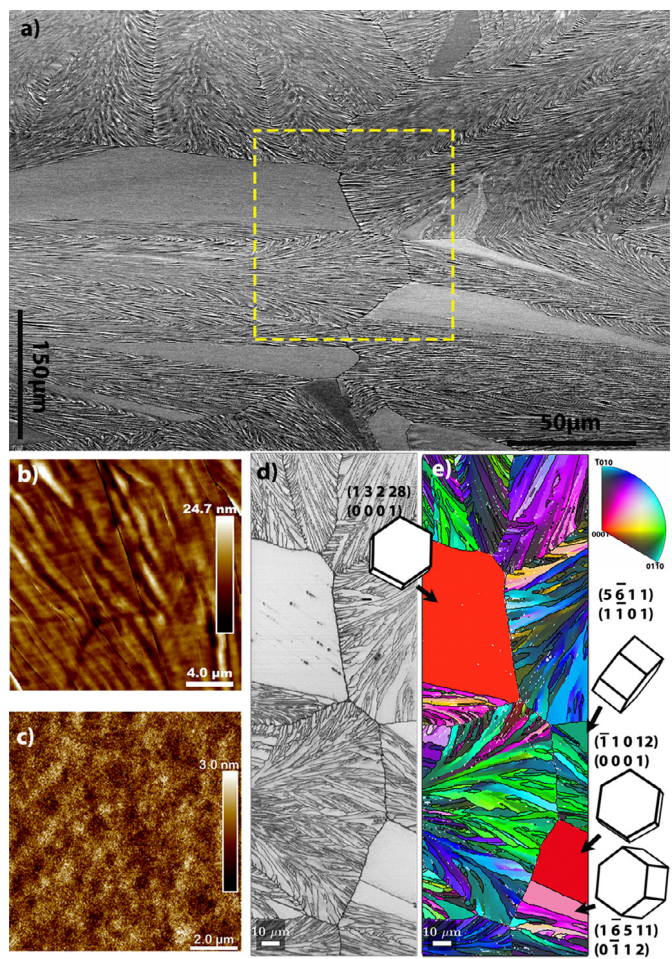

**Fig. 5.** Spherulites and single crystals coexist in GeO$_2$ on Al$_2$O$_3$ substrates annealed at 930 °C. (a) SEM image of one area where fibrous spherulites and single crystals with uniform contrast appear together. The sample is tilted by 71°. (b) and (c) are AFM images of the spherulitic area and the flat single crystals, respectively. (d) and (e) are the IQ map and [001] IPF map of the selected area inside the yellow square in (a), showing the high crystallinity of the single crystals. The unit cell sketches represent the orientation of the corresponding single crystals viewed along [001]. The hkil indices, top row: the exact orientation of the unit cells, bottom row: the simplified hkil indices close to those in the top row.

Fig. 6 shows the XRD patterns of one of the samples where spherulites and single crystals coexist. In the Grazing Incidence XRD (GIXRD) (Fig. 6(a)), the incoming X-ray is at a small incidence angle ($\omega = 0.6$ in our study) to avoid peaks from the substrates and, thus, because the film is not in specular geometry, the pattern arises from the polycrystalline spherulites which are randomly oriented. The pattern is fully consistent with the $\alpha$-quartz structure. On the other hand, the $2\theta - \omega$ (specular) scan (Fig. 6 (b)) only contains contributions from crystals planes parallel to the substrate surface, and only ($01\bar{1}2$) peak has a strong signal indicating a significant amount of crystals orient with [$01\bar{1}2$] out-of-plane and they are most likely the single crystals shown in Fig. 5. This is in agreement with the result by EBSD where [$01\bar{1}2$] is one of the dominant orientations. For the (000l) peak of quartz, the general reflection condition is l= 3n. However, the (0003) peak of GeO$_2$ quartz is very weak and generally not observed in the XRD pattern. Thus, even though (0001) may be one of the preferred orientations, it is likely that it will not show up, in agreement with our observations. Naoyuki, et al., reported a similar epitaxial relationship for SiO$_2$ quartz growth with [0001] out-of-plane orientation on (0001) Al$_2$O$_3$ sapphire substrates [33].

Fig. S9 shows the crystallization at the growth front where the crystalline phase meets the amorphous phase, and the spherulitic growth transforms into dendritic again. Moreover, a small single crystal with size of about 20 μm appears near the growth front. Interestingly, while the fibers started with a large variety of orientations, the dendrites at the growth front are also preferably oriented with the ($10\bar{1}0$) again out-of-plane, as shown in Fig. S9(c). A close look of the in-plane directions (Fig. S9(a-b)) shows that the c-axis, [0001], of the crystal lies along the long axis of the columns (the fastest growth direction), which is similar to the growth fronts with annealing temperature 880 °C shown in Fig. 4. Therefore, fibers with an original orientation different from the preferred orientation lose in the competition, stopping the growth before reaching the growth front.

Fig. S9(d-f) shows the morphology of this area in which the growth, branching of the fibers and the interaction with the waves are clearly captured, and it can be observed that these columns are continuation of the spherulitic fibers. At the columnar growth front, shown in Fig. S9(e), there are a number of small dots which produce dark contrast in the image. AFM images scanned at the growth front (Fig. S9(h)) reveal that these small dots are protuberances of about 200 nm in diameter and 10 nm in height, similar to what was observed in single nucleus crystals (Fig. 1(f) and Fig. S2), suggesting they may have similar growth mechanism. Groups of parallel lines (Fig. S9 (f) and Fig. S10) which are the slip traces [34],[35], appear. Interestingly, at the places where slip occurs, the fibers are straight, and no waves appear.

It should be noted that with the 930 °C annealing temperature, the film shows signs of melting and evaporation, with some parts of the sample in which GeO$_x$ coalesces to form droplets, revealing the bare Al$_2$O$_3$ substrate. Similarly, to crystallization, amorphization also starts from the edge (see Fig. S1 & Fig. S11). The melting point of bulk GeO$_2$ is about 1100°C [16] but our work suggests the surface energy can lower this considerably.

### 3.2. Crystallization of GeO$_2$ on MgO (110)

The crystallization on MgO substrates is shown in Fig. 7, where small circular single nucleus spherulites with dense fibers are embedded in a matrix of the huge edge-nucleated spherulites. Unlike for the growth on Al$_2$O$_3$, the entire film is successfully crystallized into quartz after annealing, which indicates that the nucleation and growth are facilitated on MgO. Fig. 7(b-d) shows the typical growth morphologies of quartz crystals on MgO which, otherwise, are similar to the morphologies on Al$_2$O$_3$: preferred nucleation at the edges with fibers growing radially and propagation of mass with wave fronts perpendicular to the growth directions, coexisting with a smaller number or isotropically grown single nucleus spherulites that nucleate in the center of the films.

### 1.%2 Crystallization of GeO$_2$ on MgAl$_2$O$_4$(110) substrates

Fig. 8(a) shows the crystallization of GeO$_2$ on MgAl$_2$O$_4$(110) substrates with annealing temperature of 830 °C. It can be observed that the films are only partially crystallized in the form of quasi-circular spherulites and a significant volume ratio of amorphous material is still present. The growth is not fully isotropic with grains with larger aspect ratio growing radially, similar to the features seen on sapphire for a higher annealing temperature of 880 °C. The image analysis shows that the central area of every spherulites has a relatively homogeneous orientation and, upon further growth, it branches into larger features with various orientations. The orientation of the different spherulites centers varies indicating no preferred orientation of nucleation, similar to what is observed on the sapphire and MgO substrates.

The morphology of such spherulites is also characterized by AFM, as shown in Fig. 8(b-c). In the center of the spherulite, the protruding core of about 500 nm in diameter serves as a nucleation center. Radial fibrils start growing from this core until it begins to branch into dendritic features to form rougher structures, being the central area of the spherulite significantly flatter than the edges. A closer view on the edge of the spherulite in (Fig. 8(c)) shows that these areas are composed of bundles of fine fibrils.

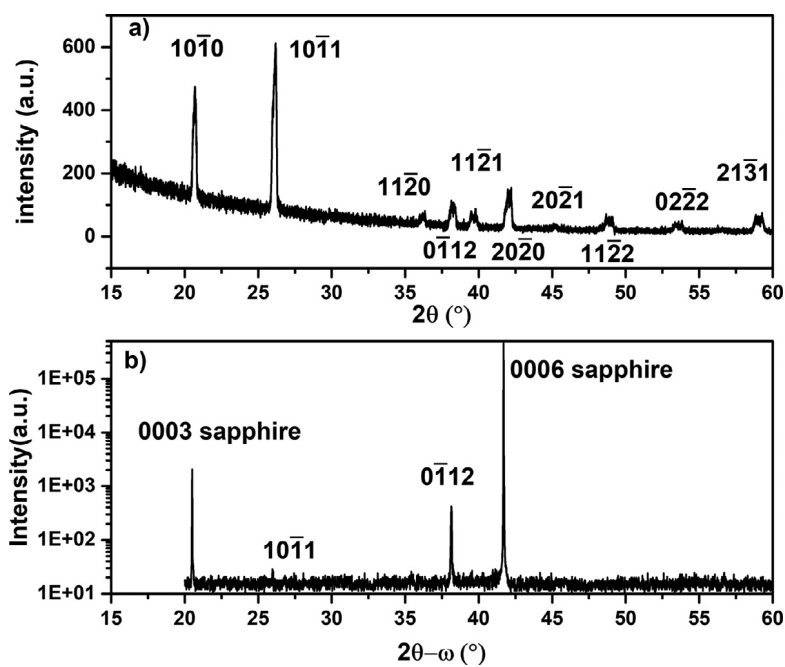

**Fig. 6.** XRD of a highly crystalline film. GIXRD scan (a) and $2\theta$-$\omega$ scan (b) of a GeO$_2$ film on Al$_2$O$_3$ substrates annealed at 930 °C. Both patterns are fully consistent with the quartz phase. The GIXRD scan shows diffraction from the randomly oriented grains; while the $2\theta$-$\omega$ scan only shows the grains that are parallel to the substrate, indicating that [01$\bar{1}$2] is the dominant out-of-plane orientation for the single crystals.

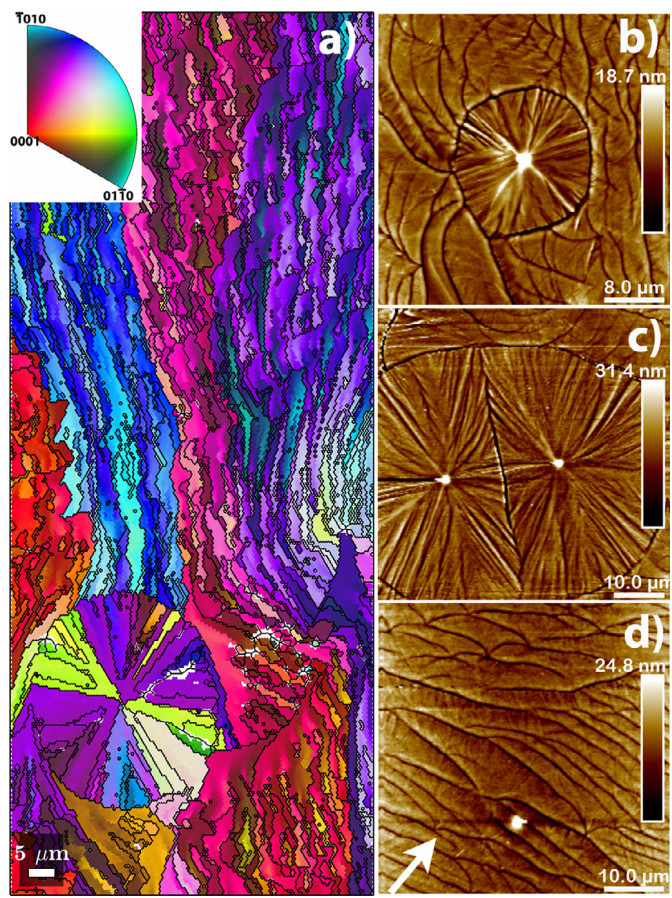

**Fig. 7.** Complete crystallization of GeO$_2$ on MgO substrates annealed at 830 °C: (a) [100] IPF Map of an area with full crystallization showing one circular spherulite embedded in the matrix of wave-like spherulites starting from the sample edge. (b-d) AFM images of different areas of the sample. White arrow in (d) points to the growth direction.

When the annealing temperature is increased to 880 °C, the number of spherulites increases significantly (see Fig. S12) and the morphology changes, as shown in Fig. 8(d-e): the spherulite boundaries are not smooth anymore and the growth front shows clear fractal growth where sub-individual dendrites grow out of the parent branches with an angle of 60°. Interestingly, crystal rods with length about 1 μm and width about 200 nm appear at the amorphous area. The closer to the spherulite, the higher they are. Concomitant with the increase in height of these rods, the neighboring trenches also become deeper, which suggests they may provide the material for the formation of the rods. Traces of these rods are also found in thin films annealed at 830 °C, although they are less noticeable.

From all the substrates we have used, only on MgAl$_2$O$_4$ there is no preferred nucleation at edge of the sample, and continuous spherulites are not observed. As can be observed from Fig. S12, a cluster of spherulites appear very close to the sample edge and on the contrary, there is only one spherulite with its nucleus exactly on the edge. Moreover, the features of the rods are also only observed on MgAl$_2$O$_4$. This suggests the nucleation and growth process on MgAl$_2$O$_4$ happens through a different mechanism.

### 3.3. Crystallization of GeO$_2$ on LAO and STO (perovskite) substrates and the formation of SrGe$_{3.3}$O$_{5.6}$

For films grown on SrTiO$_3$(STO) and Nb-doped SrTiO$_3$(Nb-STO) substrates, with an annealing temperature 830 °C, the morphology and crystallinity are almost the same. It is easier to do EBSD and obtain better resolution of electron microscopy on Nb-STO, due to its conductance. The first feature to be noticed is that, already under the optical microscope, the thin films are covered completely by colorful crystals (see Fig. S13). A SEM image in Fig. 9(a) shows these crystals, of about 200 μm in diameter, have a typical spherulitic morphology. To one's surprise, EBSD shows the film is composed of long fibers (see Fig. S13), and it does not correspond to the morphology of the spherulites in Fig. 9. Further investigation reveals that the spherulites in Fig. 9(a) are not made of quartz but a germanium oxide with Sr in it, with the quartz layer crystallized on top of it. Because GeO₂ quartz and amorphous GeOₓ are both transparent, the optical microscope only shows the germanium strontium oxide below the quartz, as observed in Fig. 9(a). Fig. S13 clearly shows the difference between the quartz GeO₂ and amorphous GeOₓ on top of the germanium oxide with Sr.

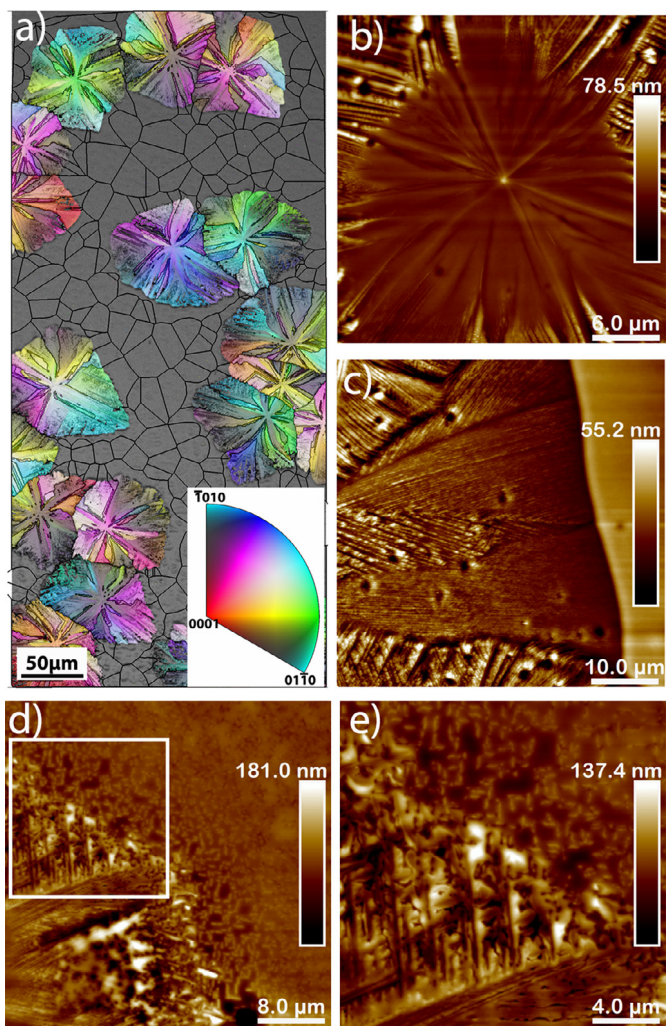

**Fig. 8.** Quartz spherulites on MgAl₂O₄ substrates annealed at 830 °C: (a-c) and 880 °C: (d-e). (a) [001] IPF map of quartz single-nucleus spherulites surrounded by amorphous GeO2 (grey area). (b) AFM scan of the central part of a spherulite. (c) AFM scan of the bundles of fine fibers at the peripheries and smooth boundary. (d) AFM image of the boundary shows a large number of rods formed in the amorphous area. (e) Blow-up of the white square in (d).

As shown in the XRD pattern in Fig. S14, besides the set of diffraction peaks characteristic of quartz GeO₂, additional peaks marked by the star symbol may be attributed to these spherulites. To understand what is going on, a cross-section of the sample was studied by Transmission Electron Microscopy (TEM) as shown in Fig. 9(c). These experiments show that the situation is somewhat more complex: from the elemental analysis, we learn that the Sr atoms of the STO substrates diffuse and react with the GeOₓ deposited on top, forming a new compound. Using Energy Dispersive X-ray Spectroscopy (EDS), we estimate the chemical formula for this new compound to be SrGe₃.₃O₅.₆, which, to the best of our knowledge, has never been reported before. With the Sr diffusing out of the STO lattice, it leaves behind protrusions of $TiO_2$ and Ti-rich STO. The $TiO_2$ protrusions are in the anatase phase and have a perfect epitaxial relationship with the STO substrate. Some Ge also diffuses into the $TiO_2$ and forms Ge-doped $TiO_2$ regions. Since Sr is a network modifier used in the glass industry [14] and it is also used in thin film as catalysis to help crystallizing quartz [5], it is of no surprise that it will react with $GeO_x$ at this temperature. On the contrary, according to Dietzel's field strength theory [14], small cations with high charge favor the formation of glass, and Ti is classified as "intermediate" (not a modifier). Only part of the $GeO_x$ layer is transformed into $SrGe_{3.3}O_{5.6}$, which suggests there could be a blocking effect by the $TiO_2$ layer for the supply of the Sr into the $GeO_x$ layer (see Fig. S15).

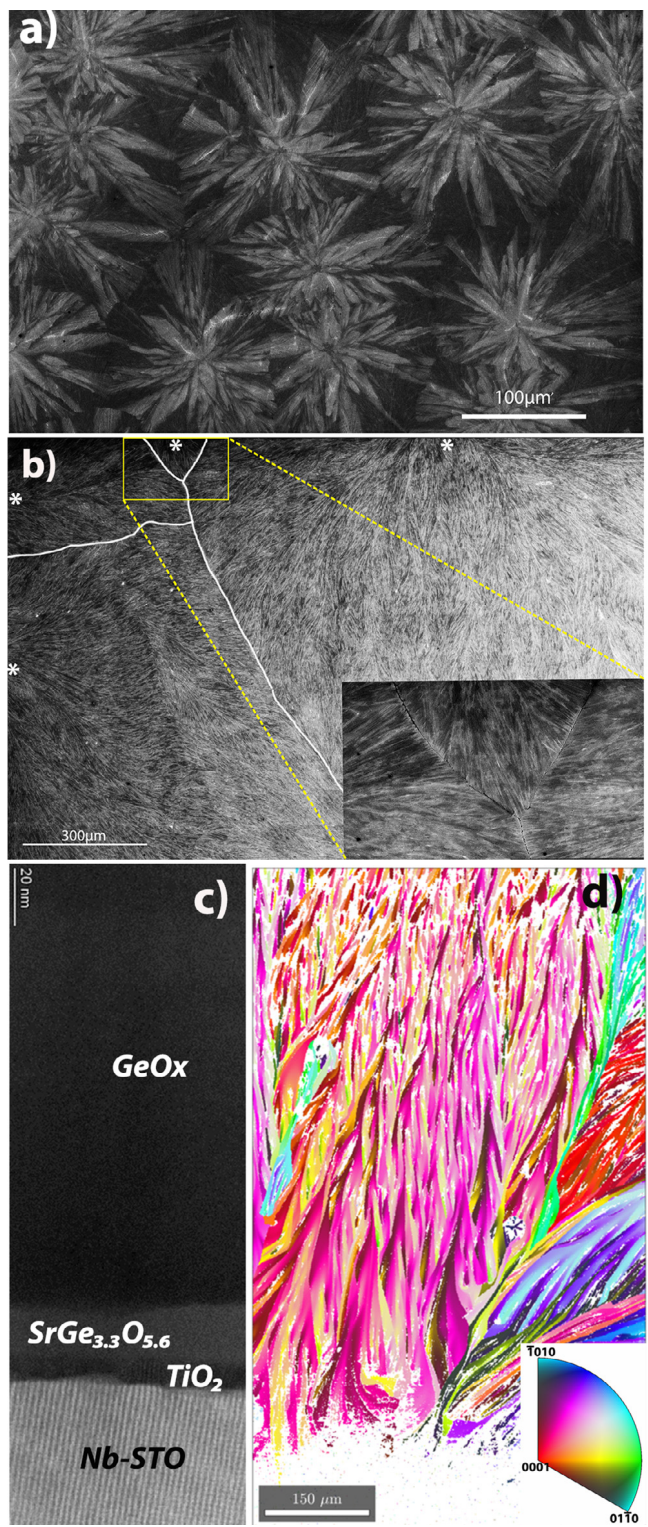

**Fig. 9.** (a) SEM image of GeO₂ on Nb-STO without using TiO₂ buffer layer. It shows the under layer of groups of spherulites which turn out to be SrGe₃.₃O₅.₆. The top layer quartz is not clearly visible due to lower contrast. (b) SEM image of GeO₂ on Nb-STO with a TiO₂ buffer layer shows the film is composed of huge quartz spherulites with nuclei on the edge of the sample. White start marks the nuclei at the edge of the sample and white lines represent the boundaries between the spherulites. The inset is the blow-up for the yellow rectangle. (c) HAADF-STEM image of the cross section of GeOx on Nb-STO without using TiO₂ buffer layer showing the complexity of the film due to the interfacial reaction between GeOx and Sr, resulting in the formation of a germanate. (d) IPF of GeO₂ grow directly on LAO without TiO₂ viewed along [010] direction. All the samples here are annealed at 830 °C.

To test this hypothesis, another layer of $TiO_2$ was deposited on Nb-STO as a buffer layer. RHEED shows the $TiO_2$ layer has grown epitaxially on STO substrates (see Fig. S16). After the growth of the buffer layer, a layer of $GeO_2$ was grown with the same recipe used for the films grown directly on the substrates. For films grown on Nb-STO with $TiO_2$ buffer layer, a direct inspection under the optical microscope shows that the colorful spherulites are gone, as shown in Fig. S17 and the entire film is crystallized. However, from the XRD pattern, a trace of the $SrGe_{3.3}O_{5.6}$ can still be found.

Fig. 9(b) shows the morphology of the film of $GeO_2$ on $TiO_2$-buffered Nb-STO substrate. Interestingly, the whole film is composed of several huge spherulites with nuclei located at the edge of the sample, as marked by the white stars, and all the fibers can be tracked back to one of these nuclei. It is clear that the fibers grow radially and keep branching until they meet the fibers from the neighboring spherulites. The white lines in Fig. 9(b) represent the boundaries between the spherulites. It can be seen that the small spherulite in the yellow rectangle lost in the competition of growth, and there is no room for its further growth, similar to what is observed on other substrates. However, unlike on other substrates, single-nucleus spherulites are not observed at the sample center. Thus, we can conclude that the growth rate on $TiO_2$-buffered STO is accelerated: once one nucleus is formed at the edge, it grows so quickly that a large part of the surface is covered before other nuclei can form.

To study the influence of $TiO_2$, thin films of $GeO_2$ were also deposited on another perovskite substrate, i.e., LAO, with and without $TiO_2$-buffer layer. For both cases, the film is fully crystallized using the same annealing temperature of STO and XRD patterns are also similar (Fig. S14), except that the sample without $TiO_2$-buffer layer has small extra peaks, marked by black squares, which could not be assigned to quartz. Both films show similar spherulitic growth. Under the optical microscope, the morphology of the thin films on LAO with buffer layer is similar to the thin films on STO with buffer layer (Fig. S17). Fig. 9(d) is the IPF Map of such a $GeO_2$ film directly grown on a LAO substrate which shows typical spherulitic growth.

## 4. Discussion

Although observations and modelling of spherulites have been reported extensively, the underlying mechanisms are still under investigation. Spherulitic-to-single crystalline transformation upon decreasing undercooling (increasing temperature) is a general trend for spherulites. Similar observation has also been found on 1,3,5-Tri-a-Naphthylbenzene [36], hippuric acid, and tetraphenyl lead spherulites [37]. One possible origin for this transformation in our films could be the different ratios of rotational $D_{rot}$ to translational $D_{tr}$ diffusion coefficients at different temperatures [38]. When the temperature is high, both $D_{rot}$ and $D_{tr}$ are large, and the material attached to the growth front is able to align itself to follow the orientation of the growth front. If upon decreasing temperature, $D_{rot}$ decreases more sharply than $D_{tr}$, the material attached on the growth front may be kinetically trapped in a relative minimum, misoriented with respect to the growth front. For the spherulites such as those in Fig. 1(d), the edges show dendritic crystallization. Transitions from spherulitic to dendritic growth are expected with increasing temperature or/and decreasing supersaturation [38]. For example, Magill and Plazek showed that when the single crystal of 1,3,5-Tri-a-Naphthylbenzene is down-quenched to low temperature, its growth front starts to branch into small fibers and finally takes on spherulitic morphology. On the contrary, if this spherulite is up-quenched back to high temperature, the growth is shifted back to single-crystalline growth [36]. So one explanation for our observations could be that the latent heat released during the crystallization increases the local temperature resulting in proper crystallization and dendritic growth at larger distances from the nucleation core.

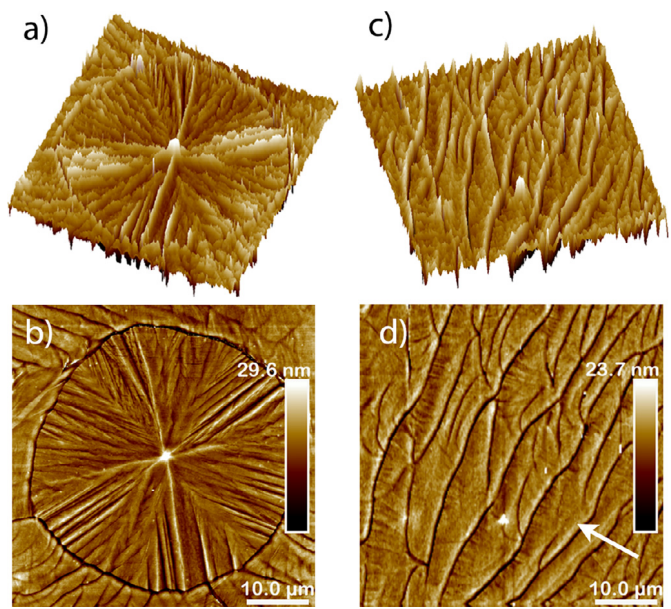

**Fig. 10.** Quartz spherulites on MgO substrate with annealing temperature 830 °C. (a)3-dimensional view of AFM image of (b) shows a typical quartz single-nucleus spherulites: radial fibrous growth starting from the core. Note that the boundary of the domain is surrounded by a deep pitch, then followed by a high wall. (c) 3-dimensional view of AFM image of (d) shows the semi-periodic structure of quartz. Generally, each wave is composed of a relatively flat area with fibers, followed by a deep pitch and a high wall (depletion and the accumulation of material). The growth direction is shown by the arrow.

For all the films, circular waves of alternating lower and higher topographic features are observed on the continuous spherulites. Similar periodic structures caused by accumulation and depletion of material during crystallization have been observed during crystallization of amorphous thin films of small molecules [40] and polymers [41,42]. As explained with Fig. 3, the superposition of various of these fronts seems to indicate that, in our case, they are rather slips systems (or perhaps cracks) formed after crystallization. However, some contribution of mass transport cannot be ruled out.

In the mass transport scenario a possible explanation is as follows: The glass temperature of $GeO_2$ is about 700 °C and our annealing temperature is significantly above this [39]. Thus, we can assume the thin film is a supercooled liquid that is able to flow during annealing. Mass diffusion results in the accumulation (dendrites) at the growth front and depletion regions (around the dendrites), as it can be clearly observed in Fig. 4(f). For the spherulitic growth, mass transport is not as prominent as during columnar growth, but it can still be present. As shown in Fig. 10 (a-b), first, all the fiber growth ends at a trench (mass depletion), then the trenches are surrounded by a wall of material (mass accumulation). Moreover, on spherulites surrounded by amorphous $GeO_x$, such as in Fig. 3(d), Fig. 8(c), height profiles show that material accumulates at the boundaries of spherulites resulting in a peak in height (Fig. S18).

At the continuous quartz area, such as shown in Fig. 10, more clearly seen in the 3-dimensional view, the accumulation of the material forms high walls (the growth direction is shown by the arrow in the Fig. 10(d)) and afterwards, fibers grow with flatter morphology. The presence of the trench suggests the depletion of the material. With the accumulation of the material, another period begins, and so on. Since the spherulitic growth is comparatively slow, it is likely that while the growth front is still propagating into the amorphous area, the crystallized part at the edge can recrystallize, which makes the whole process more complex.

These modulations have not been found on single-nucleus spherulites. As shown in Fig. S19, which images two single-nucleus spherulites at the center of the sample, with the original circular boundaries still visible, some of the fibers grow out of it forming continuous quartz, similarly to the spherulites at the edge (see Fig. S5).

## 5. Conclusions

We have successfully crystallized pure $GeO_2$ thin films into the quartz structure, on various substrates. The best crystals with sizes of hundreds of microns, were obtained on sapphire substrates. Generally speaking, they share similar behaviors of spherulitic-to-crystalline growth: at relatively low annealing temperatures, the crystallization is spherulitic, and with increasing temperatures, the growth transforms to dendritic growth at the growth front and, finally, to single crystalline growth. Lattice rotation, by which the lattice is rotating along the growth direction is observed on the spherulitic fibers and also the dendritic growth fibers but not on the single crystals.

Except for the thin films on $MgAl_2O_4$, nucleation is facilitated at the edge of the sample where half-circular spherulites with the nuclei at the sample edge are promoted, with the fibers from these spherulites growing together towards the center of the sample resulting in the crystallizing wave front sweeping from the sample edge to the center.

While on $MgAl_2O_4$ substrates, no preferential edge nucleation is found, and rods-like small crystallites form at the amorphous area before attaching to the growing spherulites, and continuous spherulites are not observed on these samples, which suggests the growth mechanism may be different than on other substrates.

For the thin films grown on STO, we found the Sr can react with the $GeO_x$ and form a new compound $SrGeO_{3.3}O_{5.6}$, which is also spherulitic, at the interface between the substrate and the layer of crystallized quartz. A buffer layer of $TiO_2$ can prevent this reaction, and the whole film is crystallized into several macro domains, suggesting the growth rate is significantly accelerated compared to other substrates. Thin film of $GeO_2$ are also crystallized on LAO with and without $TiO_2$ buffer layer, and the results are similar to those of on $TiO_2$-buffered STO.

## Declaration of Competing Interest

The authors declare that they have no known competing financial interests or personal relationships that could have appeared to influence the work reported in this paper

## Acknowledgments

This work is part of the research programme TOP-PUNT grant with project number 718.016002, which is financed by the Dutch Research Council (NWO).

We are grateful to Adrian Carretero-Genevier, Gertjan Koster, Guus Rijnders and Kit de Hond for estimulating discussions. The authors also thank Di Zhu for the experimental work on STO/LAO substrates.

## Supplementary materials

Supplementary material associated with this article can be found, in the online version, at doi:10.1016/j.actamat.2021.117069.